\documentclass[pre,amsmath,amssymb,twocolumn,showpacs,superscriptaddress,longbibliography]{revtex4-1}%,longbibliography
\usepackage{amsfonts,amsmath,amssymb,bm}
\usepackage{graphicx,graphics,float}
\usepackage{bm}
\usepackage{amssymb}
\usepackage{colordvi}
\usepackage{graphicx}
\usepackage{color}
\usepackage[colorlinks=true,linkcolor=blue,citecolor=blue,urlcolor=blue]{hyperref}%
\usepackage{hyperref}
\usepackage{comment}
\usepackage{harpoon}

\newcommand{\vep}{\varepsilon}

\newcommand{\be}{\begin{equation}}
\newcommand{\ee}{\end{equation}}
\newcommand{\bea}{\begin{eqnarray}}
\newcommand{\eea}{\end{eqnarray}}

\newcommand{\ave}[1]{\langle #1\rangle}

\def\ket#1{\vert #1 \rangle}

\begin{document}

\title{Coulomb thermoelectric drag in four-terminal mesoscopic quantum transport}

\author{Mengmeng Xi}
\affiliation{Hefei National Laboratory for Physical Sciences at Microscale, University of Science and Technology of China, 96 Jinzhai road, Hefei, Anhui, 230026, China}

\author{Rongqian Wang}
\affiliation{School of physical science and technology \&
Collaborative Innovation Center of Suzhou Nano Science and Technology, Soochow University, Suzhou 215006, China.}

\author{Jincheng Lu}\email{jincheng.lu1993@gmail.com}
\affiliation{School of physical science and technology \&
Collaborative Innovation Center of Suzhou Nano Science and Technology, Soochow University, Suzhou 215006, China.}
\address{Center for Phononics and Thermal Energy Science, China-EU Joint Lab on Nanophononics, Shanghai Key Laboratory of Special Artificial Microstructure Materials and Technology, School of Physics Science and Engineering, Tongji University, Shanghai 200092 China}

\author{Tengyun Chen}
\affiliation{Hefei National Laboratory for Physical Sciences at Microscale, University of Science and Technology of China, 96 Jinzhai road, Hefei, Anhui, 230026, China}

\author{Jian-Hua Jiang}\email{jianhuajiang@suda.edu.cn}
\affiliation{School of physical science and technology \&
Collaborative Innovation Center of Suzhou Nano Science and Technology, Soochow University, Suzhou 215006, China.}

\date{\today}
\begin{abstract}
We show that the Coulomb interaction between two circuits separated by an
insulating layer leads to unconventional thermoelectric effects, such as the cooling by thermal current effect, the transverse thermoelectric effect and Maxwell's demon effect. The first refers to cooling in one circuit induced by the thermal current in the other circuit. The middle represents electric power generation in one circuit by the temperature gradient in the other circuit. The physical picture of Coulomb drag between the two circuits is first demonstrated for the case with one quantum dot in each circuits and then elaborated for the case with two quantum
dots in each circuits. In the latter case, the heat exchange
between the two circuits can vanish. Last, we also show that the Maxwell's demon effect can be realized in the four-terminal quantum dot thermoelectric system, in which the quantum system absorbs the heat from the high-temperature heat bath and releases the same heat to the low-temperature heat bath without any energy exchange with the two heat baths. Our study reveals the role of Coulomb interaction in non-local four-terminal thermoelectric transport.
\end{abstract}

%\pacs{05.70.Ln, 84.60.-h, 88.05.De, 88.05.Bc}

\maketitle

\section{Introduction}

Thermoelectric transport is a useful tool to study the fundamental properties of quasiparticles in mesoscopic systems~\cite{gchen2005book}. Abundant information of quasiparticle dynamics and fluctuations, phase
coherence and statistics can be revealed in thermoelectric
transport~\cite{DubiRMP,JiangCRP,BENENTI20171}. Most existing researches are focused on elastic
thermoelectric transport such as ballistic transport and resonant
tunneling in mesoscopic systems~\cite{Sivan,Mahan,GoldsmidBook,natureVenka,RMP06,Jiang06JAP,LinkePRL,ZhouJunPRL,MyJAP}.
Only recently, there has been a surge
of interest in studying thermoelectric transport involving inelastic
processes where electron energy is not conserved during the transport.
It was found that inelastic thermoelectric devices can provide higher
efficiency and larger output power than conventional elastic
thermoelectric devices made of the same material~\cite{David2011PRB,Rafael,DavidPRL,Lena2012,Jiang2012,Jordan2013,Nanotechnology,JiangSR,JiangBijayPRB17,Rongqian,Jiang2017,JiangNearfield,MyPRBDiode}.
Furthermore, several unconventional effects, such as the cooling by heating effect~\cite{Cooling1,Cooling2,Cooling3,MyPRBTransistor} and the linear-response thermal transistor effect~\cite{bli2004prl,bli2006apl,Jiangtransistors,Transistor9,Transistor1}, have been discovered in inelastic thermoelectric devices. It was also proved that those effects cannot take place in elastic thermoelectric device.

Several types of inelastic thermoelectric devices have been studied,
involving, e.g., electron-phonon~\cite{OraPRB2010,RenPRB12,Jiang2013,BijayJiang} or electron-electron interaction~\cite{CoulombRMP,SanchezPRL10,RenPRB12,HartmannPRL,ZhangPRE15,thierCRP,transistor-yang,SanchezPRRes,PRB20,tabatabaei}.
In general, electron-phonon interaction is more important at high temperatures, whereas electron-electron interactions are important at very low temperatures. In this work we shall show the inelastic thermoelectric transport at the low-temperature regime and discuss the key role of the electron-electron interaction on it.

It has been shown that in a three-terminal device with one quantum
dot (QD), electronic noises (essentially heat energy) from the third
terminal can induce electrical power generation for the source and
drain terminals~\cite{Rafael,Sanchez13NJP}. The underlying mechanism is the inelastic Coulomb
scattering which induces a directional transport when the left-right
inversion symmetry is broken as in other quantum rachets.
In those studies, heat in the third electronic terminal is harvested to produce
electrical energy across the source and the drain. Recently, Whitney
{\sl et al.} found that in a four-terminal QD thermoelectric device,
electrical energy across the source and the drain can be produced by
harvesting heat from two additional capacitively coupled
terminals~\cite{WhitneyPhysE}. Remarkably, the produced electrical energy is finite when
the total heat injected into the QD system vanishes.

In this work, we show that in a four-terminal mesoscopic system
cooling of the source can be achieved by passing a perpendicular
heat current between the two additional terminals, even if these two
terminals do not exchange charge with the source or the drain. The
cooling by heat current effect is driven by the Coulomb drag
between electrons in the two circuits even though they are separated
by a charge insulation layer. This effect may happen in a number of
four-terminal mesoscopic thermoelectric systems when the Coulomb
interaction between two circuits is strong. Here we focus on
illustration of the effect via correlated transport through Coulomb
coupled QDs with particular spatial symmetries. These symmetries are
realized by engineering the energy dependent coupling between the QDs
and the four terminals. With a simple two QDs set-up, the cooling by
heat current effect can be realized. In a set-up with two pairs of
double QDs, the cooling by heat current effect can be realized with
the total heat exchanged between the two circuits vanishes. Such a
limit is consistent with the thermodynamic analysis that the cooling
of the source is driven by the heat flow in the other circuit from the
hot terminal to the cold terminal, instead of the total heat flux
exchanged between the two circuits.

\section{Four-terminal quantum-dots thermoelectric systems}

\subsection{Double quantum-dots thermoelectric systems}

\begin{figure}[htb]
\includegraphics[width=7.0cm]{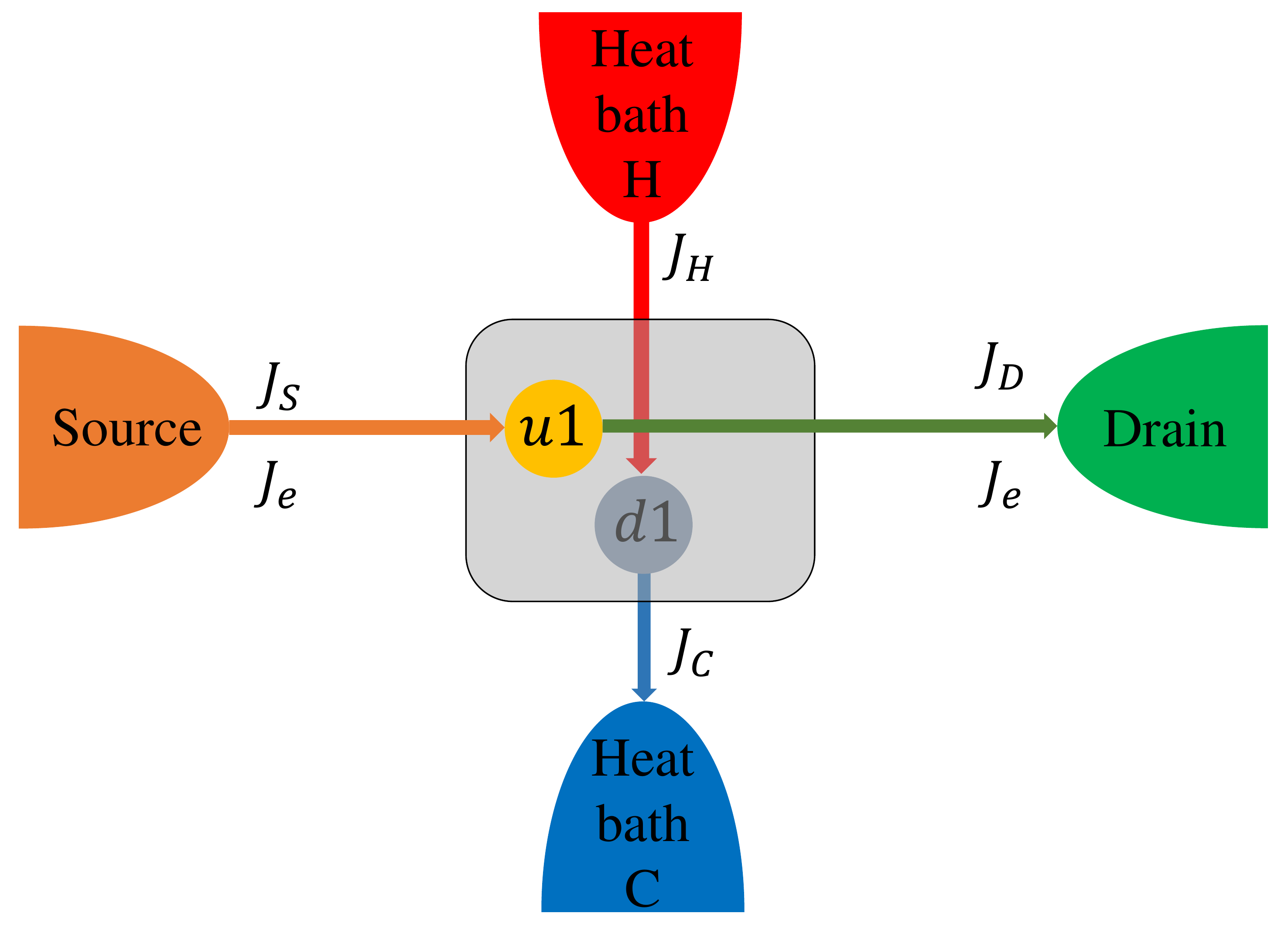}
\caption{Schematic of four-terminal QDs thermoelectric system. Beside the source and the drain, there are two other electrodes, $H$ and $C$. Only the heat currents flowing out/into these electrodes are considered, since their electrochemical potentials are set to the equilibrium value. In experiments, these two electrodes (termed as ``heat baths'') and the QD $d1$ are fabricated in a lower layer. In the upper layer, there are the source, the drain, and the QD $u1$. The lower and upper layers are separated by an insulating layer (depicted as the gray shadow). Electrons in the two QDs interact with each other via Coulomb (repulsive) interaction.}~\label{fig:fig1}
\end{figure}

We first consider a four-terminal double QDs thermoelectric system
which is illustrated in Fig.~\ref{fig:fig1}. The system
can be regarded as two independent circuits without Coulomb interaction. The source, the drain and
the upper quantum dot form a circuit, while the other two electrodes
and the lower quantum dot form another circuit. The two circuits are
separated by an insulating layer. Thermoelectric in the two circuits
are correlated once the interaction between electrons in the two QDs
are considered. In general, Coulomb interaction between the two QDs
can cause drag effect and inelastic thermoelectric transport in the two
circuits, which leads to unconventional thermoelectric phenomena. To
demonstrate the underlying physics, we shall consider the Coulomb
interaction of the following form,
\be
H_{\rm Coulomb} = E_C n_1 n_2 ,
\ee
where $n_1$ and $n_2$ are the number of electrons in the two QDs,
respectively. $E_C$ is the strength of Coulomb
interaction. We shall show in this work that such a simple
interaction leads to novel thermoelectric phenomena in four-terminal
systems.

Beside the source and the drain, there are other two electrodes
(electronic reservoirs). In our set-up, their electrochemical
potentials are set to the equilibrium value so that they serve as
``heat baths'', i.e., charge motions between them do not produce or
consume any electrical energy. We thus consider only the heat current
flowing out of the $H$ reservoir, $J_H$, and the heat current flowing
into the $C$ reservoir, $J_C$ (we suppose $T_H>T_C$,
where $T_H$ and $T_C$ are the temperature of the heat bath $H$ and $C$, respectively.). In contrast, the
source and the drain can have potentials differing from the equilibrium value,
leading to charge motion under (self-consistent) electrical fields. We
hence consider both the charge current between the source and the
drain, $J_e$, and the heat current flowing out of (into) the source
(drain) terminal, $J_S$ ($J_D$).

Although there are four heat currents, energy conservation impose the
following restriction,
\be
J_S + \frac{\mu_S}{e} J_e + J_H = J_C + J_D + \frac{\mu_D}{e}J_e ,
\ee
where $\mu_S$ and $\mu_D$ are the electrochemical potentials of the
source and the drain, respectively. It is useful to
consider the following combinations
\be
J_{in} = J_H - J_C, \quad J_q = \frac{1}{2}(J_H+J_C) .
\ee
Here $J_{in}$ can be understood as the total heat injected into the
QDs system from the two heat baths $H$ and $C$. $J_q$ is regarded
as the heat flow (exchange) between the reservoirs $H$ and $C$.
We shall choose the independent heat currents as three of the four
heat currents, e.g., $J_S$, $J_{in}$ and $J_q$. The affinities
corresponding to the three heat currents and the electrical current
can be found by analyzing the total entropy production~\cite{JiangPRE,JiangPRL},
\begin{subequations}
\begin{align}
& \partial_t S_{\rm tot} = \sum_i J_i A_i, \\
& A_e = \frac{\mu_S-\mu_D}{eT_D}, \quad A_S \equiv \frac{1}{T_D}
-\frac{1}{T_S}, \\
& A_{in} \equiv \frac{1}{T_D} -
\frac{1}{2T_H} - \frac{1}{2T_C}, \quad  A_q \equiv
\frac{1}{T_C}-\frac{1}{T_H} .
\end{align}
\end{subequations}

To demonstrate the underlying physics, in the following we shall
consider the simple situation where there is only a single energy level that
allows tunneling between each electrode to the connected QD.

\begin{figure}[htb]
\includegraphics[width=7.0cm]{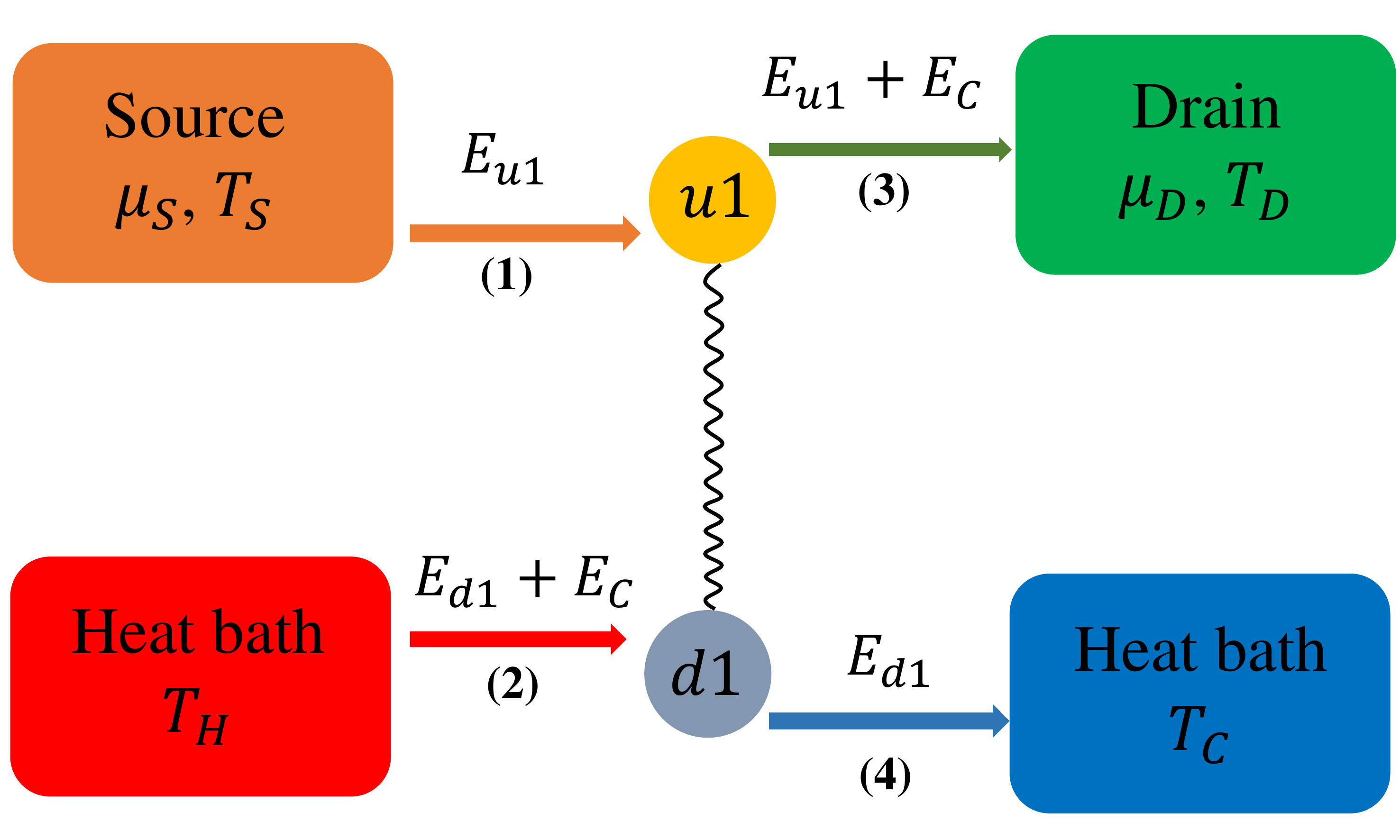}
\caption{Illustration of correlated electron transport through the QDs $u1$ and $d1$. An electron tunnels between the source and the QD $u1$ only at energy $E_{u1}$ (i.e., when there is no electron in the QD $d1$). The electron tunnels between the drain and the QD $u1$ can only take place at energy $E_{u1}+E_C$ (i.e., with an electron in the QD $d1$). For the QD $d1$, electron tunnels between the heat bath $H$ and the QD only at energy $E_{d1}+E_C$ (i.e., when there is an electron in the QD $u1$). The electron tunnels between the QD $d1$ and the heat bath $C$ only at energy $E_{d1}$. The interaction between electron, depicted as the wavy lines, is the crucial element for the correlated inelastic transport through the QDs $u1$ and $d1$. For the electron motion indicated by the arrows, an energy $E_C$ is pumped into the source and drain subsystem from the $H$ and $C$ subsystem.}~\label{fig:u1d1}
\end{figure}

We now present the microscopic theory for thermoelectric transport in
the QDs system. Explicitly we shall construct the expressions for the
electrical and heat currents as functions of the electrochemical
potentials, $\mu_S$ and $\mu_D$, as well as the temperatures, $T_S$,
$T_D$, $T_H$ and $T_C$. As shown in Fig.~\ref{fig:u1d1},
we consider particular configurations where
QDs $u1$ and $d1$ are capacitively coupled, whereas
there is no other coupling between QD $u1$ and other QDs. We also
assume that the intradot Coulomb repulsion is so strong that double
occupancy is forbidden. In such configurations, the energy for
electron in the QDs are given by
\begin{subequations}
\begin{align}
& E_{u1,n_{d1}} = E_{u1} + n_{d1} E_C, \\
& E_{d1,n_{u1}} = E_{d1} + n_{u1} E_C, \\
& \quad n_{u1}, n_{d1}=0,1 .
\end{align}
\end{subequations}
Here $E_{u1}$ and $E_{d1}$ are the electron energies for QDs $u1$ and
$d1$ if there is no interaction between QDs. $n_{u1}$ and $n_{d1}$
denote the number of electrons in the $u1$ and $d1$ QDs,
respectively.

We now study the steady-state transport through $u1$ and $d1$ using
the master equation method as in Ref.~\cite{Rafael}. There are
four quantum states for electrons in the two QDs $u1$ and $d1$, respectively,
characterized by the quantum numbers $n_{u1}$ and $n_{d1}$, and
denoted as $\ket{n_{u1},n_{d1}}$. The electron density matrix of the
QDs system at the steady-state is characterized by the four-component
vector $\rho=(\rho_{0,0}, \rho_{1,0}, \rho_{0,1}, \rho_{1,1})^T$. The
steady-state transport currents are determined by the steady-state
density matrix $\rho$, which is obtained by solving the
master equation for steady-state, $\partial_t
\rho=\hat{L}_{(1)}\rho=0$. Here

\begin{widetext}
\be
\hat{L}_{(1)} = \left( \begin{array}{cccc}
    - k_{u1,0}^+ - k_{d1,0}^+ & k_{u1,0}^- & k_{d1,0}^- &
    0 \\
    k_{u1,0}^+ & - k_{u1,0}^- -k_{d1,1}^+ & 0 &
    k_{d1,1}^- \\
    k_{d1,0}^+ & 0 & - k_{d1,0}^- -k_{u1,1}^+ &
    k_{u1,1}^- \\
    0 & k_{d1,1}^+ & k_{u1,1}^+ & - k_{d1,1}^- -  k_{u1,1}^- \\
    \end{array}\right)
\ee
\end{widetext}
The transition rates are
\begin{subequations}
\begin{align}
& k_{u1,n_{d1}}^\pm \equiv k_{S,u1, n_{d1}}^\pm + k_{D,u1, n_{d1}}^\pm, \\
& k_{d1,n_{u1}}^\pm \equiv k_{H,d1, n_{u1}}^\pm + k_{C,d1, n_{u1}}^\pm,
\end{align}
\end{subequations}
where the superscript $+$ ($-$) representing the rates for the
processes with electron entering into (leaving) the QD. For the ``$k$''
in the right hand sides of the equations, the first two subscripts
denote the reservoir and the QD as the starting or end point of the
tunneling, while the third subscript denotes the number of electrons
in the capacitively coupled QD (e.g., for tunneling into the QD
$u1$, the capacitively coupled QD is $d1$). And
\begin{subequations}
\begin{align}
& k_{i,u1, n_{d1}}^+\equiv \Gamma_{i,u1, n_{d1}}f\left(\frac{E_{u1,n_{d1}}-\mu_i}{k_BT_i}\right),
\ \ i=S,D, \\
& k_{i,d1, n_{u1}}^+\equiv \Gamma_{i,d1,n_{u1}} f\left(\frac{E_{d1,n_{u1}}}{k_BT_i}\right), \
\ i=H,C,\\
& k_{i,u1,n_{d1}}^- \equiv \Gamma_{i,u1,n_{d1}} - k_{i,u1,n_{d1}}^+,  \ \ i=S, D, \\
& k_{i,d1,n_{u1}}^- \equiv \Gamma_{i,d1,n_{u1}} - k_{i,d1,n_{u1}}^+,  \ \ i=H, C ,
\end{align}
\end{subequations}
where $\Gamma_{i,u1, n_{d1}}$ is the tunneling rate between the
reservoir $i$ and the QD $u1$ at energy $E_{u1,n_{d1}}$ (the other
tunneling rates are defined similarly), and $f(x)=1/(e^x+1)$ is the
Fermi-Dirac distribution function.

In the above equations, we have set
\be
\mu_H=\mu_C\equiv0
\ee
as both the equilibrium electrochemical potential and the energy zero for
electrons. The tunneling rates can, in principle, be calculated via
the Fermi golden rule~\cite{Jiang2012,Jiangtransistors}. In this work, however, we assume that they can
be tuned at will in experiments (e.g., by changing the channel width
of the leads, or attaching QDs that are strongly coupled with the
leads; in the former we change the step-function like tunneling rates,
while in the latter Lorentzian shape tunneling rates can be
formed)~\cite{prete,jaliel-exper}. It has been found that asymmetry in the tunneling rates
$\Gamma_{i}$ (as functions of energy) is crucial to induce
inelastic thermoelectric transport in the QDs
system~\cite{Rafael,SothmannPRB}.

\subsection{Four quantum-dots thermoelectric system}

\begin{figure}[htb]
\includegraphics[width=7cm]{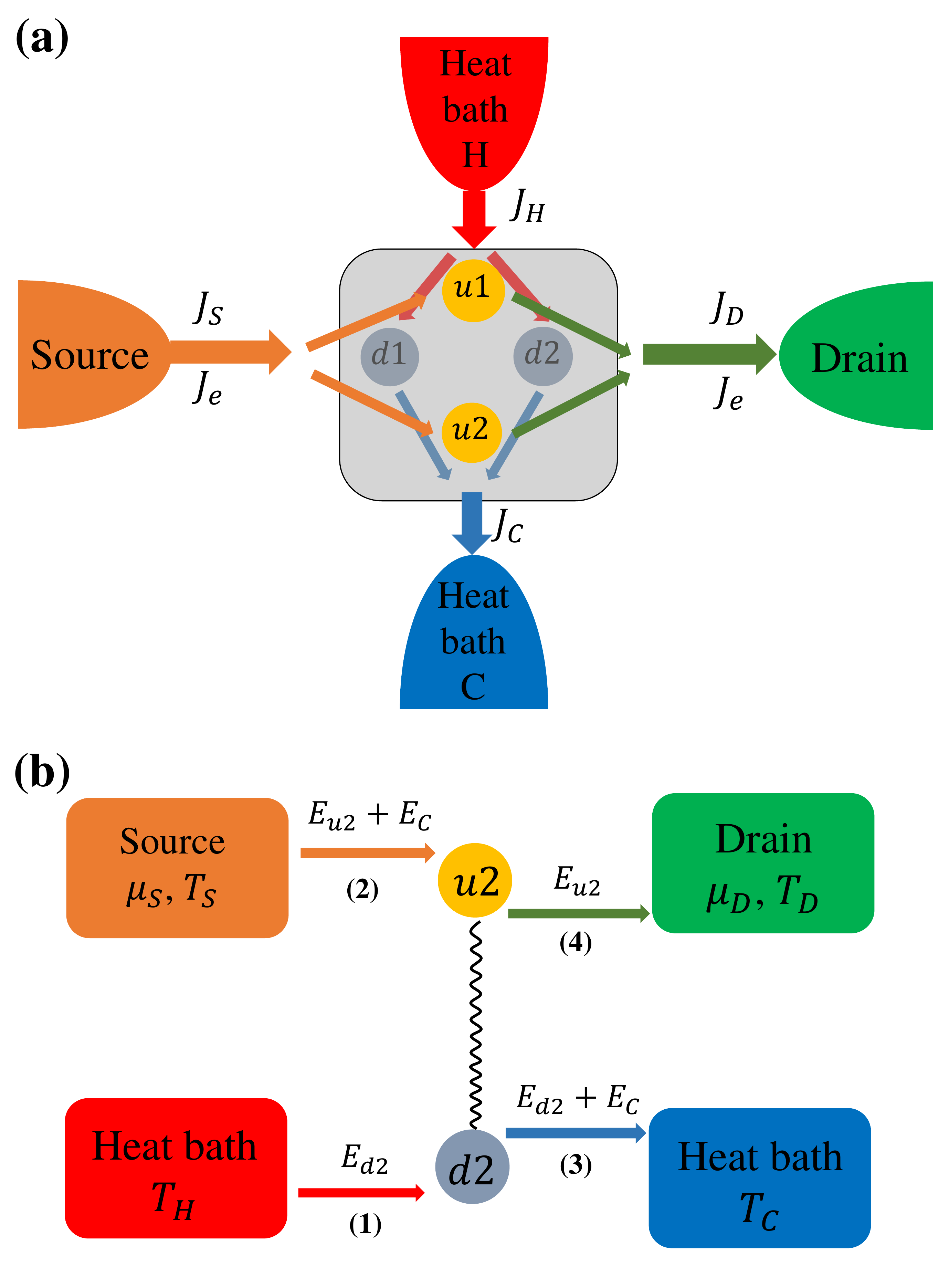}
\caption{(a) Schematic of four-terminal QDs thermoelectric system. Beside the source and the drain, there are two other electrodes. Only the heat currents flowing out/in these electrodes are considered, since their electrochemical potentials are set to the equilibrium value. In experiments, these two electrodes (termed as ``heat baths'') and the two QDs $d1$ and $d2$ can be fabricated in a lower layer. The source and drain electrodes together with the QDs $u1$ and $u2$ can be fabricated in the upper layer. The lower layer and the upper layer are separated by an insulating layer (depicted as the gray shadow). But electrons in the QDs can interact with each other via the long range Coulomb interaction. (b) Correlated electron tunneling through the QDs $u2$ and $d2$. Similar to electron tunneling through the $u1$ and $d1$ QDs, but with different energy configuration for electron tunneling. For electron motion indicated by the arrows in the figure, an energy $E_C$ is provided by the source and drain subsystem to the $H$ and $C$ subsystem. The arrows indicate one possible direction of electron motion, which can be reversed, depending on the voltage and temperature biases. The numbers in the brackets indicate the sequences of the correlated tunneling processes.}~\label{fig:4QDsystem}
\end{figure}

In order to realize the cooling by heat current effect through the Coulomb interaction, we consider the four-terminal four QDs thermoelectric system, as shown in Fig.~\ref{fig:4QDsystem}(a). In this particular configurations,  QDs $u2$ and $d2$ are capacitively coupled, whereas there is no other coupling between QD $u2$ and other QDs [see Fig.~\ref{fig:4QDsystem}(b)].

Similarly, the expressions hold for QDs $u2$ and $d2$ are
\begin{align}
& E_{u2,n_{d2}} = E_{u2} + n_{d2} E_C, \\
& E_{d2,n_{u2}} = E_{d2} + n_{u2} E_C, \\
& \quad n_{u2}, n_{d2}=0, 1 .
\end{align}
For simplicity, we have assumed that the charge interaction strength between
the QDs $u2$ and $d2$ are the same as that for the QDs $u1$ and $d1$.
Transport through QDs system can actually be divided into two parts:
the four-terminal transport through the pair of QDs $u1$ and $d1$ and
that through the pair of QDs $u2$ and $d2$. These two channels are
{\em independent} of each other.

Transport equations through the other pair of QDs, $u2$ and $d2$, can be solved similarly.
In fact, the matrix $\hat{L}_{(2)}$ is the same as $\hat{L}_{(1)}$ with only $u1$ replaced
by $u2$ and $d1$ replaced by $d2$. We shall use the vector
$\rho^\prime=(\rho^\prime_{0,0}, \rho^\prime_{1,0}, \rho^\prime_{0,1},
\rho^\prime_{1,1})^T$ to represent the density matrix of electrons in
the QDs $u2$ and $d2$ at the transport steady-state. Here each element
of the density matrix $\rho^\prime_{n_{u2}n_{d2}}$ carries the index
characterizing the quantum states of electrons
in the QDs $u2$ and $d2$, $\ket{n_{u2},n_{d2}}$.

After numerically solving the master equations at the steady-state transport~\cite{Rafael}
\be
\hat{L}_{(1)}\rho=0,\quad \hat{L}_{(2)}\rho^\prime=0,
\ee
one can determine the steady-state transport currents from $\rho$
and $\rho^\prime$ as,
\begin{widetext}
\begin{subequations}
\begin{align}
& J_{e} = \sum_{n_{d1}} e(k_{S,u1,n_{d1}}^+\rho_{0,n_{d1}} -
k_{S,u1,n_{d1}}^-\rho_{1,n_{d1}} ) +
\sum_{n_{d2}}e(k_{S,u2,n_{d2}}^+\rho^\prime_{0,n_{d2}} -
k_{S,u2,n_{d2}}^-\rho^\prime_{1,n_{d2}} ), \\
& J_S = \sum_{n_{d1}} (E_{u1,n_{d1}} - \mu_S) (
k_{S,u1,n_{d1}}^+\rho_{0,n_{d1}} - k_{S,u1,n_{d1}}^-\rho_{1,n_{d1}}) +
\sum_{n_{d2}} (E_{u2,n_{d2}} - \mu_S)
(k_{S,u2,n_{d2}}^+\rho^\prime_{0,n_{d2}} - k_{S,u2,n_{d2}}^- \rho^\prime_{1,n_{d2}} ) , \\
& J_H = \sum_{n_{u1}} E_{d1,n_{u1}}
(k_{H,d1,n_{u1}}^+\rho_{n_{u1},0}-k_{H,d1,n_{u1}}^-\rho_{n_{u1},1}) +
\sum_{n_{u2}} E_{d2,n_{u2}}
(k_{H,d2,n_{u2}}^+ \rho^\prime_{n_{u2},0}- k_{H,d2,n_{u2}}^- \rho^\prime_{n_{u2},1}) , \\
& J_C = \sum_{n_{u1}} E_{d1,n_{u1}} (k_{C,d1,n_{u1}}^- \rho_{n_{u1},1} -
k_{C,d1,n_{u1}}^+ \rho_{n_{u1},0} ) + \sum_{n_{u2}} E_{d2,n_{u2}} (k_{C,d2,n_{u2}}^- \rho^\prime_{n_{u2},1} -
k_{C,d2,n_{u2}}^+ \rho^\prime_{n_{u2},0} ) .
\end{align}
\end{subequations}
\end{widetext}
We can obtain $J_{in}$ and $J_q$ from the two currents, $J_H$ and $J_C$.

\subsection{Inelastic thermoelectric transport through correlated tunneling}

We now study a special configuration that allows unconventional
thermoelectric transport through drag effects. In this configuration,
we choose the following tunneling rates,
%\begin{widetext}
\begin{equation}
\begin{aligned}
& \Gamma_{S,u1,n_{d1}=0} = \Gamma_0, \quad \Gamma_{S,u1,n_{d1}=1} = 0 , \\
& \Gamma_{D,u1,n_{d1}=0} = 0, \quad \Gamma_{D,u1,n_{d1}=1} = \Gamma_0 , \\
& \Gamma_{H,d1,n_{u1}=0} = 0, \quad \Gamma_{H,d1,n_{u1}=1} = \Gamma_0 , \\
& \Gamma_{C,d1,n_{u1}=0} = \Gamma_0, \quad \Gamma_{C,d1,n_{u1}=1} = 0 ; \\
& \Gamma_{S,u2,n_{d2}=0} = 0 , \quad \Gamma_{S,u2,n_{d2}=1} = \Gamma_0, \\
& \Gamma_{D,u2,n_{d2}=0} = \Gamma_0, \quad \Gamma_{D,u2,n_{d2}=1} = 0 , \\
& \Gamma_{H,d2,n_{u2}=0} = \Gamma_0, \quad \Gamma_{H,d2,n_{u2}=1} = 0, \\
& \Gamma_{C,d2,n_{u2}=0} = 0, \quad \Gamma_{C,d2,n_{u2}=1} = \Gamma_0,
\end{aligned}
\label{config}
\end{equation}
%\end{widetext}
where $\Gamma_0$ is a constant for the tunneling rates. The above
configuration is schematically illustrated in Fig.~\ref{fig:u1d1} and Fig.~\ref{fig:4QDsystem}(b). For such a
configuration, in the QDs $u1$ and $d1$ only the following process
of electron motion (and its time-reversal) is possible,

\begin{equation}
\begin{aligned}
& \ket{0,0}\xrightarrow[QD\ u1:\
E_{u1}]{\Gamma_{S,u1,0}}\ket{1,0}\xrightarrow[QD\ d1:\
E_{d1}+E_C]{\Gamma_{H,d1,1}}\ket{1,1}\\
& \xrightarrow[QD\ u1:\
E_{u1}+E_C]{\Gamma_{D,u1,1}}\ket{0,1}\xrightarrow[QD\ d1:\ E_{d1}]{\Gamma_{C,d1,0}}\ket{0,0} .
\end{aligned}
\end{equation}

The above expression represents the correlated electron tunneling through the QDs $u1$ and $d1$. The quantum states are labeled by the two integers $n_{u1}, n_{d1}$. In the above expression, below each arrow is the QD of which electron is tunneling into or outside and the energy of the
electron that tunnels, while above each arrow is the tunneling
rate. The consequence of the above correlated tunneling processes is
an electron traverses from the $H$ reservoir to the $C$ reservoir via
the QD $d1$, in the meanwhile an electron traverses from the source to
the drain via the QD $u1$. The interaction between the electrons in
the two QDs provides the mechanism for passing energy between the two
channels. The electron traverse from $H$ to $C$ loses energy $E_C$,
while the electron traverses from the source to the drain and gains energy $E_C$. The above correlated tunneling processes, as the only passage
for electron transport through the QDs $u1$ and $d1$, also dictates
the direction correlation between the electron flowing in the two
channels: when an electron goes from $H$ to $C$, there has to be an
electron moving from the source to the drain, otherwise the energy is
not conserved! Such locking between electron motions in the two
channels is the key for the unconventional thermoelectric transport in our system.

Similarly, for transport through the QDs $u2$ and $d2$,
only the following process and its time-reversal is allowed,
\begin{equation}
\begin{aligned}
& \ket{0,0}\xrightarrow[QD\ d2:\ E_{d2}]{\Gamma_{H,d2,0}}\ket{0,1}\xrightarrow[QD\ u2:\
E_{u2}+E_C]{\Gamma_{S,u2,1}}\ket{1,1}\\
& \xrightarrow[QD\ d2:\
E_{d2}+E_C]{\Gamma_{C,d2,1}}\ket{1,0}\xrightarrow[QD\ u2:\ E_{u2}]{\Gamma_{D,u2,0}}\ket{0,0} .
\end{aligned}
\end{equation}
Each time the above process (the reversed process) takes place, an
electron traverse from $H$ to $C$, and another electron traverse from
the source to the drain. There are always an equal number of electrons
traveling from $H$ to $C$ as the number of electrons traveling from
the source to the drain.

According to the above picture, the electrical and heat currents can be
expressed as
\begin{subequations}
\begin{align}
& J_e = e \sum I_i, \\
& J_S = E_{u1} I_1 + (E_{u2}+E_C) I_2, \\
& J_H = (E_{d1}+E_C) I_1 + E_{d2} I_2 ,\\
& J_C = E_{d1} I_1 + (E_{d2}+E_C) I_2 ,\\
& J_{in} = E_C (I_1 - I_2), \\
& J_q = (E_{d1} + \frac{1}{2}E_C) I_1 + (E_{d2} +
\frac{1}{2} E_C) I_2 , \label{rel-I}
\end{align}
\end{subequations}
where
\begin{subequations}
\begin{align}
& I_1 = \gamma_1^{-1} (k^+_{u1,0} k^+_{d1,1} k^-_{u1,1} k^-_{d1,0} -
k^+_{d1,0} k^+_{u1,1} k^-_{d1,1} k^-_{u1,0}) , \\
& I_2 = \gamma_2^{-1} (k^+_{d2,0}k^+_{u2,1}k^-_{d2,1}k^-_{u2,0} -
k^+_{u2,0}k^+_{d2,1}k^-_{u2,1}k^-_{d2,0}) ,
\end{align}
\end{subequations}
are the electron number current through the QD $u1$ (or $d1$), and
the electron number current through the QD $u2$ (or $d2$),
respectively. According to Ref.~\cite{Rafael},
\begin{equation}
\begin{aligned}
\gamma_1 = &\sum_{\alpha=u1,d1}\sum_{i=\pm} \sum_{n=0,1}
k^{-i}_{\alpha, n} (k^i_{\alpha,1-n}k^i_{-\alpha,n} \\
&+ k^{-i}_{-\alpha,n}\sum_j k^j_{-\alpha,n}\delta_{|1+i|,2n}) , \\
\gamma_1 = &\sum_{\alpha=u2,d2}\sum_{i=\pm} \sum_{n=0,1}
k^{-i}_{\alpha, n} (k^i_{\alpha,1-n}k^i_{-\alpha,n} \\
&+ k^{-i}_{-\alpha,n}\sum_j k^j_{-\alpha,n}\delta_{|1+i|,2n}) ,
\end{aligned}
\end{equation}
where $-u1=d1$ and $-u2=d2$. From the relations between currents we
can also obtain the relations between the linear-response transport
coefficients. If $I_i = \sum_j M_{ij} A_j$ ($i,j=e,S,in,q$) where
$M_{ij}$ are the linear-response transport coefficients, then
\begin{widetext}
\begin{subequations}
\begin{align}
& M_{e,e} = T G, \quad M_{e,S} = e^{-1}TG \left[p_1 E_{u1} + p_2 (E_{u2}+E_C) \right], \quad
M_{e,in}=e^{-1}TG E_C (p_1 - p_2), \\
& M_{e,q} = e^{-1}TG  \left[ p_1 E_{d1} + p_2 E_{d2} + \frac{1}{2}E_C\right]
, \quad M_{S,S} = e^{-2} TG  \left[p_1 E_{u1}^2 + p_2 (E_{u2}+E_C)^2 \right],\\
& M_{S,in} = e^{-2}TG\left[p_1 E_{u1}E_C - p_2 (E_{u2}+E_C)E_C \right], \\
& M_{S,q} = e^{-2}TG \left[p_1 E_{u1} (E_{d1} + \frac{1}{2}E_C) + p_2
(E_{u2}+E_C)(E_{d2} + \frac{1}{2}E_C) \right] ,\\
& M_{in,in} = e^{-2}TG E_C^2, \quad M_{in,q} =  e^{-2}TG\left[p_1
E_C(E_{d1} + \frac{1}{2}E_C) - p_2 E_C (E_{d2} + \frac{1}{2}E_C) \right],\\
& M_{q,q} = e^{-2}TG \left[p_1 (E_{d1} + \frac{1}{2}E_C)^2 + p_2 (E_{d2} +
\frac{1}{2}E_C)^2 \right]
\end{align}
\end{subequations}
\end{widetext}
where $G$ is the total conductance, and $p_i=G_i/G$ ($i=1,2$) with
$G_i$ being the conductance for the transport through the QDs pair
$u1$ and $d1$ (if $i=1$) or that for the transport through the QDs
pair $u2$ and $d2$ (if $i=2$).

\subsection{Unconventional Thermoelectric energy conversion}
There are three different types of thermoelectric effects, associated with the following Seebeck coefficients~\cite{JiangJAP,MyJAP},
\begin{subequations}
\begin{align}
& S_S = \frac{M_{e,S}}{T^2G} = \frac{k_B}{e}\frac{\ave{a_S}}{k_BT} , \\
& S_{in} =  \frac{M_{e,in}}{T^2G} = \frac{k_B}{e}\frac{\ave{a_{in}}}{k_BT} , \\
& S_q = \frac{M_{e,q}}{T^2G} = \frac{k_B}{e}\frac{ \ave{a_q} }{k_BT} ,
\end{align}
\end{subequations}
where
\begin{subequations}
\begin{align}
& \ave{{\cal O}} = \sum_{i=1,2} p_i {\cal O}_i\\
& a_{S,1} = E_{u1}, \quad a_{S,2} = E_{u2}+E_C ,\\
& a_{in,1} = E_C, \quad a_{in,2} = - E_C,\\
& a_{q,1} = E_{d1} + \frac{1}{2}E_C, \quad a_{q,2} = E_{d2} +
\frac{1}{2}E_C .
\end{align}
\end{subequations}
In our system which breaks the left-right inversion symmetry and up-down inversion symmetry [see Eq.~(\ref{config})], if $p_1=p_2$, the Seebeck coefficient $S_{in}$ vanishes. The other two Seebeck coefficients are nonzero even at the particle-hole symmetric limit
with $E_{u1}=E_{d1}=E_{u2}=E_{d2}=0$. The figures of merit for these three thermoelectric effects can be expressed as
\begin{subequations}
\begin{align}
& Z_ST = \frac{\ave{a_S}^2}{\ave{a_S^2}-\ave{a_S}^2}, \\
& Z_{in}T = \frac{\ave{a_{in}}^2}{\ave{a_{in}^2}-\ave{a_{in}}^2} ,\\
& Z_q T = \frac{\ave{a_{q}}^2}{\ave{a_{q}^2}-\ave{a_{q}}^2} .
\end{align}
\end{subequations}
These figures of merit can be rather high when the variances of $a_{S}$, $a_{in}$, and $a_q$ are small.

\section{Cooling by thermal current}

We now study the cooling by thermal current effect, i.e., cooling the
source (i.e., $J_S>0$ even though $T_S<T_D$ and $A_S<0$) as driven by
the thermal current $J_q$ when
\begin{subequations}
\begin{align}
& \mu_S=\mu_D=0, \quad {\rm i.e.},\quad A_e=0 ,\\
& T_C = \frac{1}{\frac{2}{T_D} - \frac{1}{T_H} }, \quad {\rm i.e.},\quad
A_{in}=0 .\label{A0}
\end{align}
\end{subequations}
We consider the situations with $T_H>T_D>T_C$. $T_D$ is the reference
temperature that may be set by the substrate. This effect can take
place only when $J_qA_q>0$, i.e., the negative entropy production (i.e., positive free energy
production) during the cooling of the source is compensated by the
positive entropy production during the thermal conduction $J_q$ (i.e.,
free energy consumption, $J_qA_q>0$), so that the total entropy
production is non-negative,~\cite{kedem,JiangPRE}
\be
\partial_t S_{\rm tot} = J_q A_q + J_S A_S \ge 0 .\label{cbhstot}
\ee
In the linear-response regime, to have $J_S>0$ while $A_S<0$, one needs
a positive and large enough $M_{S,q}$.

\begin{figure}[htb]
\includegraphics[height=7.5cm]{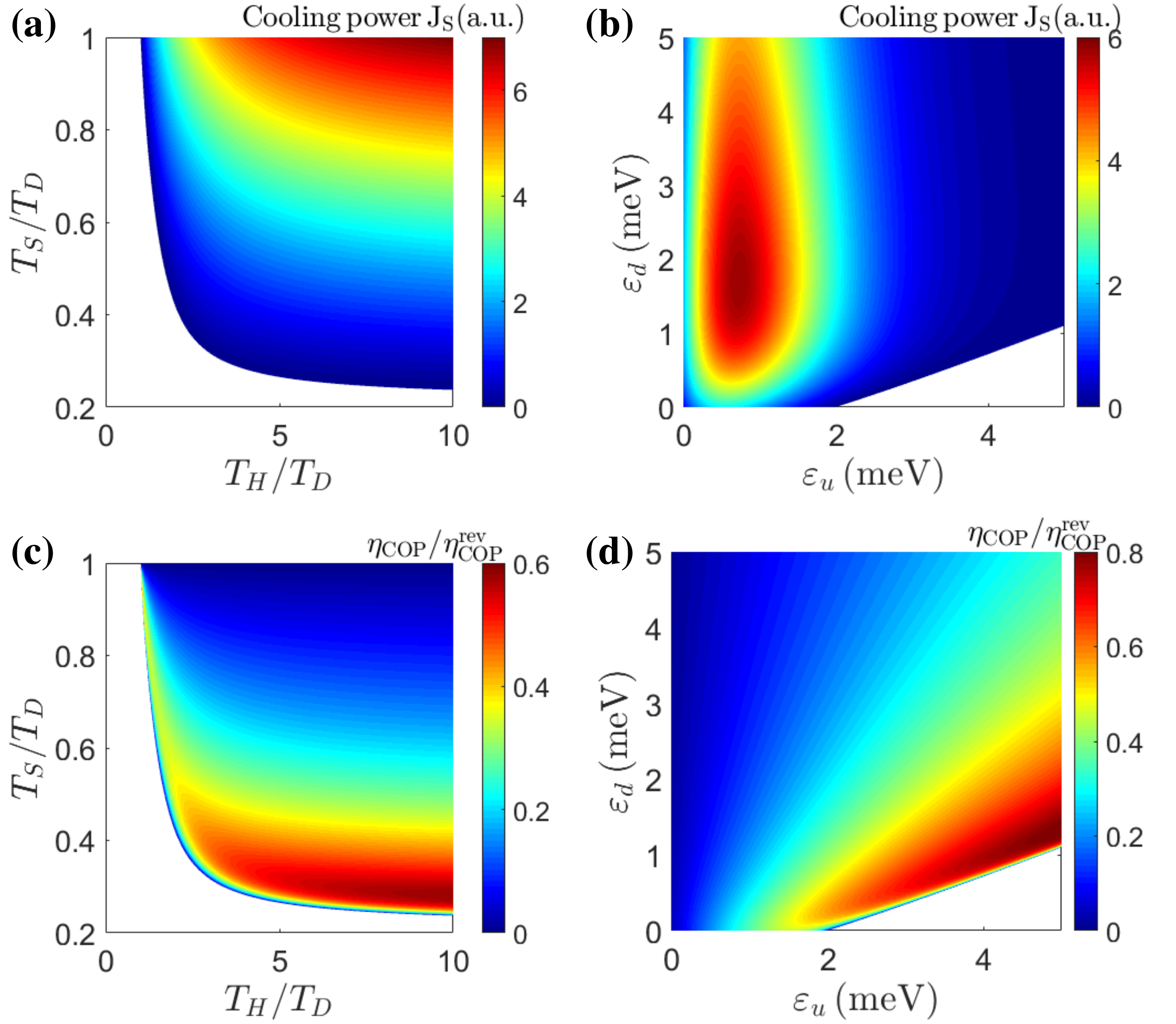}
\caption{(a) Cooling power $J_S$ as functions of temperature ratios $T_S/T_D$ and $T_H/T_D$ for $E_{u1}=E_{u2}=E_{d1}=E_{d2}=1$~meV. (b) Cooling power $J_S$ as functions of QDs energies $E_{u1}=E_{u2}=\vep_u$ and $E_{d1}=E_{d2}=\vep_d$ for $k_BT_H=6$~meV and $k_BT_S=0.6$~meV. (c) The COP ratio $\eta_{\rm COP}/\eta_{\rm COP}^{\rm rev}$ for the cooling by thermal current effect as functions of the temperature ratios $T_S/T_D$ and $T_H/T_D$ when QDs energy is the same as in (a). (d) $\eta_{\rm COP}/\eta_{\rm COP}^{\rm rev}$ as functions of QDs energies $\vep_u$ and $\vep_d$ for the same parameters as in (b). Common parameters: $\mu_S=\mu_D=0$, $k_BT_D=1$~meV, $T_C=1/(2/T_D-1/T_H)$, $\Gamma_0=0.1$~meV, and $E_C=2$~meV.}~\label{fig:THTS}
\end{figure}

The coefficient of performance (COP) for the cooling by thermal current
effect is given by the ratio of the two heat currents
\be
\eta_{\rm COP} = \frac{J_S}{J_q} = \frac{ E_{u1} I_1 + (E_{u2}+E_C) I_2 }{  (E_{d1} + \frac{1}{2}E_C) I_1 + (E_{d2} +
\frac{1}{2} E_C) I_2 } .
\ee
%The regime with $I_1=I_2$ (i.e, $J_{in}=0$~\cite{WhitneyPhysE}) is called the strong-coupling (SC) regime~\cite{kedem,Cooling2}, where one has
%\be
%\eta_{\rm COP}^{\rm SC} = \frac{ E_{u1} + E_{u2} + E_C }{ E_{d1} + E_{d2} + E_C  } .
%\ee
From Eq.~(\ref{cbhstot}) and the definition of the COP we find that
\be
\eta_{\rm COP} =
\left(T_S A_q - \frac{T_S{\partial_t}S_{\rm tot}}{J_q} \right)\frac{T_D}{T_D-T_S} .
\ee
The reversible COP is
\be
\eta^{\rm rev}_{\rm COP} = \left(\frac{T_S}{T_C} - \frac{T_S}{T_H} \right)
\frac{T_D}{T_D-T_S} = -\frac{A_q}{A_S} .
\ee
We regard $T_D$ as the reference temperature fixed by the substrate
and $T_C$ as determined by Eq.~(\ref{A0}), so that only the
temperatures $T_S$ and $T_H$ are independent variables. The working
condition for the cooling driven by the thermal current $J_q$ is
restricted by
\be
0\le \eta_{\rm COP} \le -\frac{A_q}{A_S} .
\ee
%In the strong coupling regime, one has
%\be
%0\le \frac{ E_{u1} + E_{u2} + E_C }{ E_{d1} + E_{d2} + E_C  } \le -\frac{A_q}{A_S}
%\ee
At reversible COP $\eta^{\rm rev}_{\rm COP} = -\frac{A_q}{A_S}$ the cooling
power vanishes, since the entropy production and the currents
vanish~\cite{kedem,JiangOra,JiangPRE}.

In Fig.~\ref{fig:THTS}(a) the cooling power of our nonelectric refrigerator is
large when the temperature of the source $T_S$ is close to the
temperature of drain $T_D$ and when the temperature of the heat bath
$T_H$ is much higher than the temperature of the drain, $T_H\gg T_D$.
From Fig.~\ref{fig:THTS}(a) the lowest temperature of the source $T_S$ that can be
cooled down is about $0.25T_D$. In the white region the device cannot
function as a refrigerator. As shown in Fig.~3(b), for a given
temperature of $T_H=6T_D$ and $T_S=0.6T_D$, the cooling power is large
for $E_{u1}=E_{u2}=\vep_u$ around $k_BT_D$ ($k_BT_D=1$~meV throughout
this paper), particularly when $\vep_d\gtrsim \vep_u$. In Fig.~\ref{fig:THTS}(c) the ratio of the COP over the
reversible COP $\eta_{\rm COP}/\eta_{\rm COP}^{\rm rev}$ is high when the temperature of the source $T_S$ is low. This is the regime where the cooling power is rather low. Similarly, as shown in Fig.~\ref{fig:THTS}(d), the COP
ratio $\eta_{\rm COP}/\eta_{\rm COP}^{\rm rev}$ is high when $\vep_d$ is small, where the cooling power is very low. The COP can be quite a considerable fraction of the reversible COP, indicating the coupling is strong and the dissipation is low for such optimized efficiency
conditions~\cite{JiangOra}. This phenomenon that the efficiency is optimized when the output power is low (for small dissipations), known as the efficiency-power trade-off, is consistent with the existing
literature~\cite{JiangOra,JiangPRE,trade-off}.

\begin{figure}[htb]
\includegraphics[height=3.9cm]{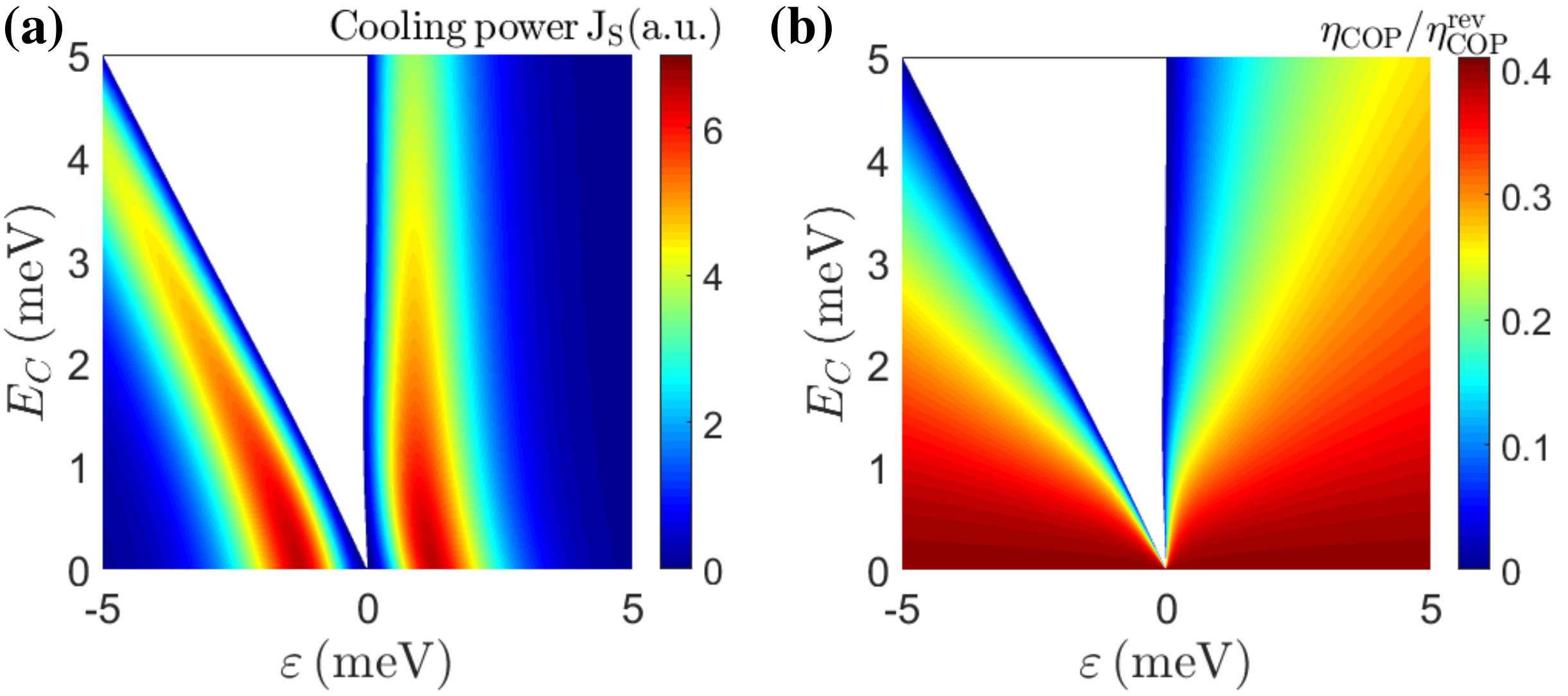}
\caption{(a) Cooling power $J_S$ and (b) COP ratio $\eta_{\rm COP}/\eta_{\rm COP}^{\rm rev}$ as functions of the QDs energies $E_{u1}=E_{u2}=E_{d1}=E_{d2}=\vep$ and $E_C$. Other parameters: $\mu_S=\mu_D=0$, $k_BT_D=1$~meV, $k_BT_H=6$~meV, $k_BT_S=0.6$~meV, $T_C=1/(2/T_D-1/T_H)$, and $\Gamma_0=0.1$~meV.}~\label{fig:epEC}
\end{figure}

We now study more on the QDs energy dependence of the cooling power $J_S$
and the COP ratio $\eta_{\rm COP}/\eta_{\rm COP}^{\rm rev}$ for several different
configurations. We first consider the case with $E_{u1}=E_{u2}=E_{d1}=E_{d2}=\vep$ and study how the energy $\vep$ and the charge interaction energy $E_C$ affect the power and the COP.
First the cooling by thermal current effect holds for all $E_C>0$ when
$\vep>0$. However, as shown in Fig.~\ref{fig:epEC}(a), for $\vep<0$, the cooling by
thermal current effect holds only when $E_C<-\vep$. This is because in
this regime, $M_{S,q} = e^{-2}TG[p_1 \vep ( \vep + \frac{1}{2}E_C) + p_2
(\vep + E_C)( \vep + \frac{1}{2}E_C) ]$. The condition for positive
$M_{S,q}$ holds when $E_C<-\vep$. From Figs.~\ref{fig:epEC}(a) and \ref{fig:epEC}(b), we note
that the optimal parameters for both COP and cooling power can be
$\vep\sim k_BT_D$ and $E_C\sim k_BT_D$, or $\vep\sim -2 k_BT_D$ and
$E_C\sim k_BT_D$ ($k_BT_D=1$~meV). Although the COP ratio increases
with decreasing $E_C$ for any given $\vep$, the limit with $E_C\to 0$ does not validate our theory.

\begin{figure}[htb]
\includegraphics[height=3.9cm]{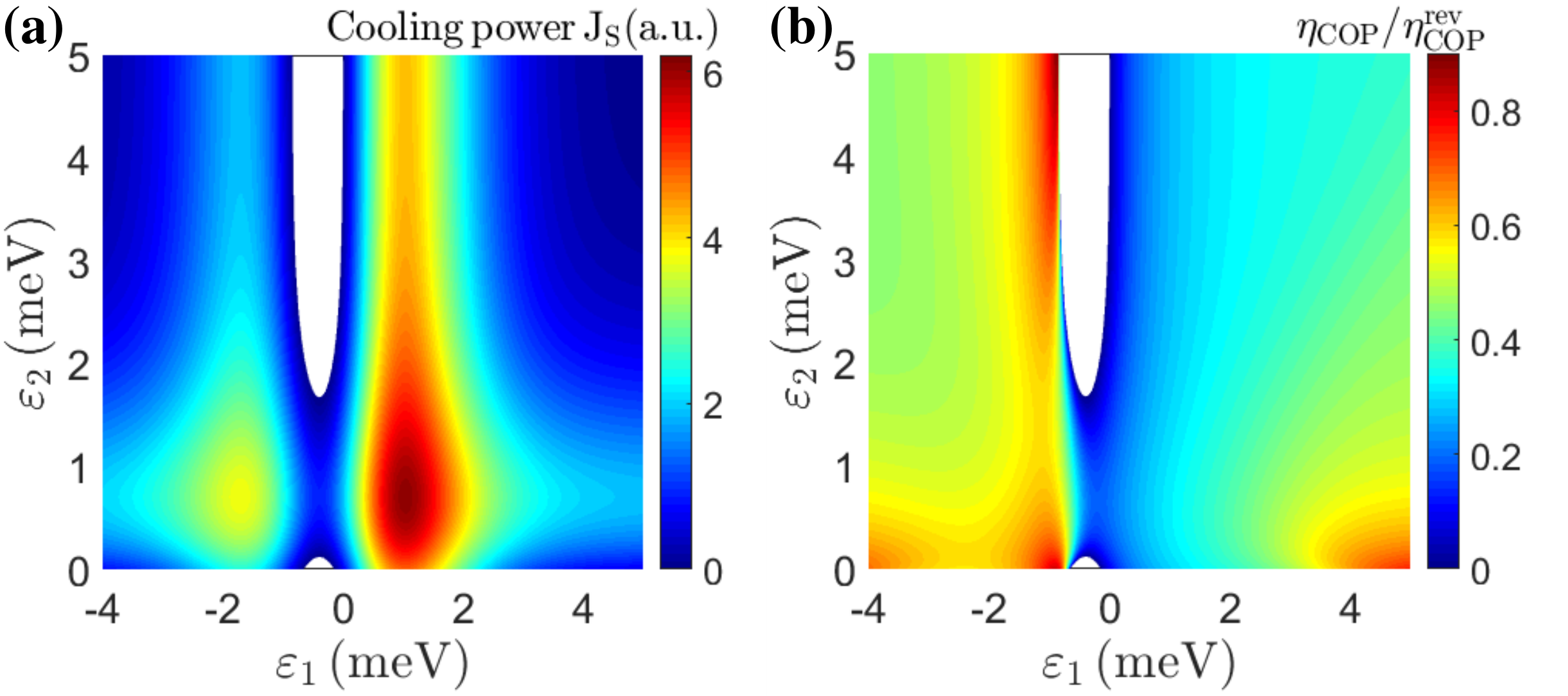}
\caption{(a) Cooling power $J_S$ and (b) COP ratio $\eta_{\rm COP}/\eta_{\rm COP}^{\rm rev}$ as functions of the QDs energies $E_{u1}=E_{d1}=\vep_1$ and $E_{u2}=E_{d2}=\vep_2$. Other parameters: $\mu_S=\mu_D=0$, $k_BT_D=1$~meV, $k_BT_H=6$~meV, $k_BT_S=0.6$~meV, $T_C=1/(2/T_D-1/T_H)$, $E_C=1$~meV, and $\Gamma_0=0.1$~meV.}~\label{fig:ep1ep2}
\end{figure}

If we allow the energy configuration to be different between the pair of QDs $u1$ and $d1$ and the other pair of QDs $u2$ and $d2$ in a way that $E_{u1}=E_{d1}=\vep_1$ and $E_{u2}=E_{d2}=\vep_2$ with $E_C=1$~meV, we can study how the cooling performance varies with the energies in the two channels, $\vep_1$ and $\vep_2$. From the Fig.~\ref{fig:ep1ep2}(a), we find that the cooling power is maximized at $\vep_1\sim \vep_2\sim 1$~meV. As shown in Fig.~\ref{fig:ep1ep2}(b), the COP ratio is rather optimized when $\vep_1\sim -1$~meV for large $\vep_2$. This observation again demonstrates the efficiency-power trade off~\cite{JiangOra,JiangPRE,trade-off}.

\section{Four-terminal thermoelectric system as a Maxwell's demon}

Maxwell's demon is a hypothetical demon, which can detect and control the motion of a single molecule. It was conceived by James Clerk Maxwell in 1871 to explain the possibility of violating the second law of thermodynamics.
The demon is assumed to control a door, which separates two boxes to realize the non-equilibrium gas distribution~\cite{Maxwellbook}. This operation that appeared to be in violation of the second law of thermodynamics, decreasing the entropy of the gas without any work input.
Despite the complexity, such Maxwell's demon has been realized in various systems in different forms~\cite{demon1,demon2,demon3,demon4}. For example, the QD multi-terminal systems can be possibly realized as a Maxwell's demon by using the inelastic transport without any measurement of individual particles or any feedback involved~\cite{SanchezPRRes,SanchezDemon,demonDQD,MyPRBdemon}.

\begin{figure}[htb]
\begin{center}
\centering\includegraphics[width=7.5cm]{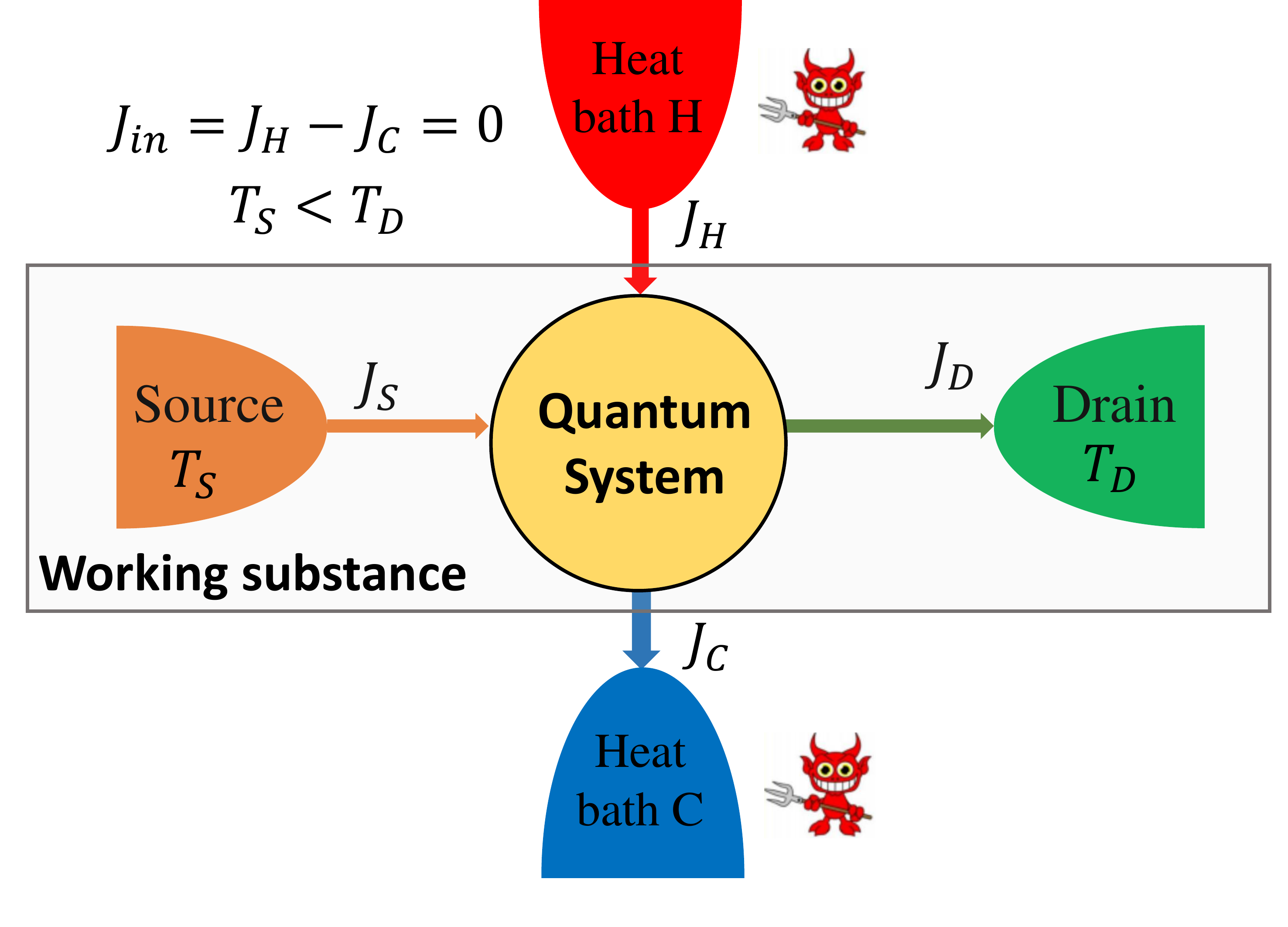}
\caption{Illustration of a four-terminal mesoscopic thermoelectric device as a Maxwell demon. The Maxwell's demon does not supply any energy to the working substance, i.e., the total heat current injected into the central quantum system from the two thermal baths is zero, $J_{in}=J_H-J_C=0$.} ~\label{fig:demon-system}
\end{center}
\end{figure}

As shown in Fig.~\ref{fig:demon-system}, here we introduce a Maxwell's demon based on two electronic thermal baths (the hot bath $H$ and cold bath $C$) which can reduce the entropy of the working substance (the source and the drain), without changing the energies and particles of the working substance. The nonequilibrium Maxwell's demon is used to reverse the natural direction of the source and drain where the drain has a higher temperature.
As we introduced in the previous section, when we realize the cooling by thermal current effect, the entropy of the source and drain is reduced, while the entropy of the whole thermoelectric system is increased. This operation seemingly break the the second law of thermodynamics, the entropy production of the two heat baths $H$ and $C$ compensates for that of the source and drain.

%%%%%%%%%%%%%%%%%%%%%%%%%%%%%%%%%%%%%%%%%%%%%%%%%%%%%%%%%%%%%%%%%%
%%%%%%%%%%%%%%%%%%%%%%%%%%%%%%%%%%%%%%%%%%%%%%%%%%%%%%%%%%%%%%%%%%

The condition at which the Maxwell's demon neither injects nor extracts heat or energy into the working substance is~\cite{MyPRBdemon}
\begin{equation}
J_{in} \,=\, 0.
\end{equation}
The power of the Maxwell demon vanishes when the temperature gradient between the two heat baths $H$ and $C$ is vanishing.

\begin{figure}[htb]
\includegraphics[height=3.9cm]{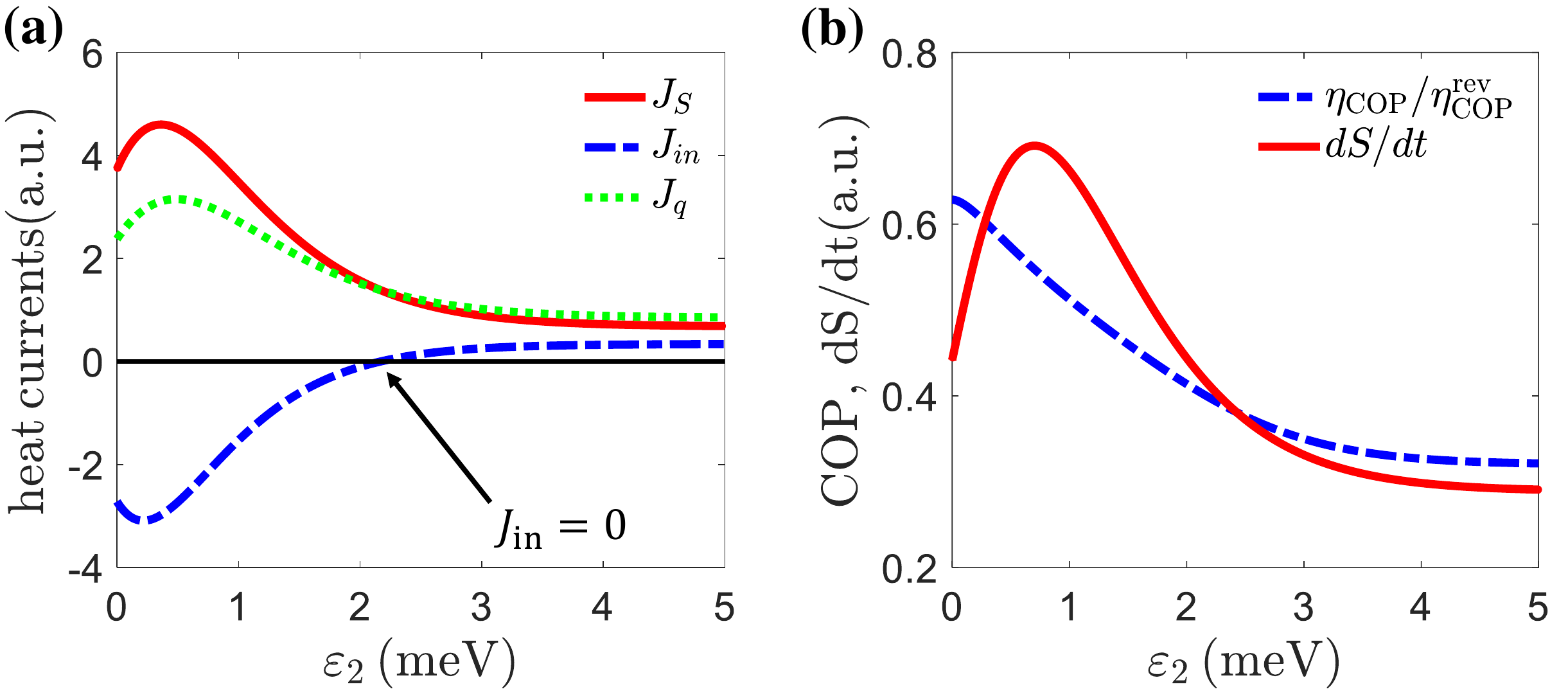}
\caption{(a) Cooling power $J_S$, the total injected heat $J_{in}$, and the thermal current $J_q$ as a function of the QD energy $E_{u2}=E_{d2}=\vep_2$. (b) COP ratio $\eta_{\rm COP}/\eta_{\rm COP}^{\rm rev}$, entropy production rate $\partial _tS_{\rm tot}$ as a function of the QDs energy $E_{u2}=E_{d2}=\varepsilon_2$. Other parameters: $\mu_S=\mu_D=0$, $k_BT_D=1$~meV, $k_BT_H=6$~meV, $k_BT_S=0.6$~meV, $T_C=1/(2/T_D-1/T_H)$, $E_{u1}=E_{d1}=\vep_1=4$~meV, $E_C=2$~meV, and $\Gamma_0=0.1$~meV.}~\label{fig:ep2}
\end{figure}

\begin{figure}[htb]
\includegraphics[height=3.9cm]{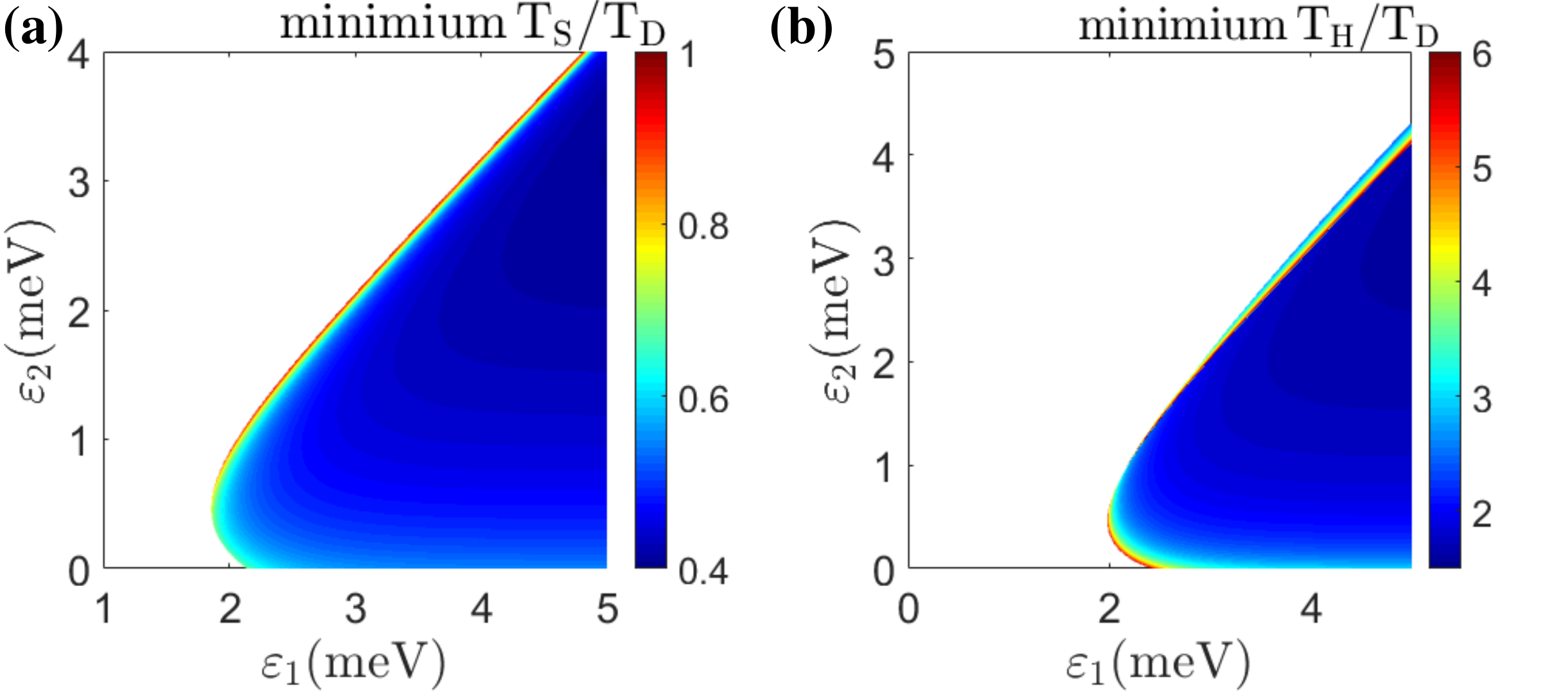}
\caption{(a) The lowest temperature of source $T_S$ that can be cooled down via the cooling by transverse current effect as functions of the QDs energies $\vep_1$ and $\vep_2$ for $k_BT_H=6$~meV and $J_{in}=0$. (b) The lowest temperature of the hot heat bath T H that can perform the cooling by transverse current effect as functions of the QDs energies $\vep_1$ and $\vep_2$ for $k_BT_S=0.6$~meV and $J_{in}=0$.}~\label{fig:minTHTS}
\end{figure}

\begin{figure}[htb]
\includegraphics[height=3.9cm]{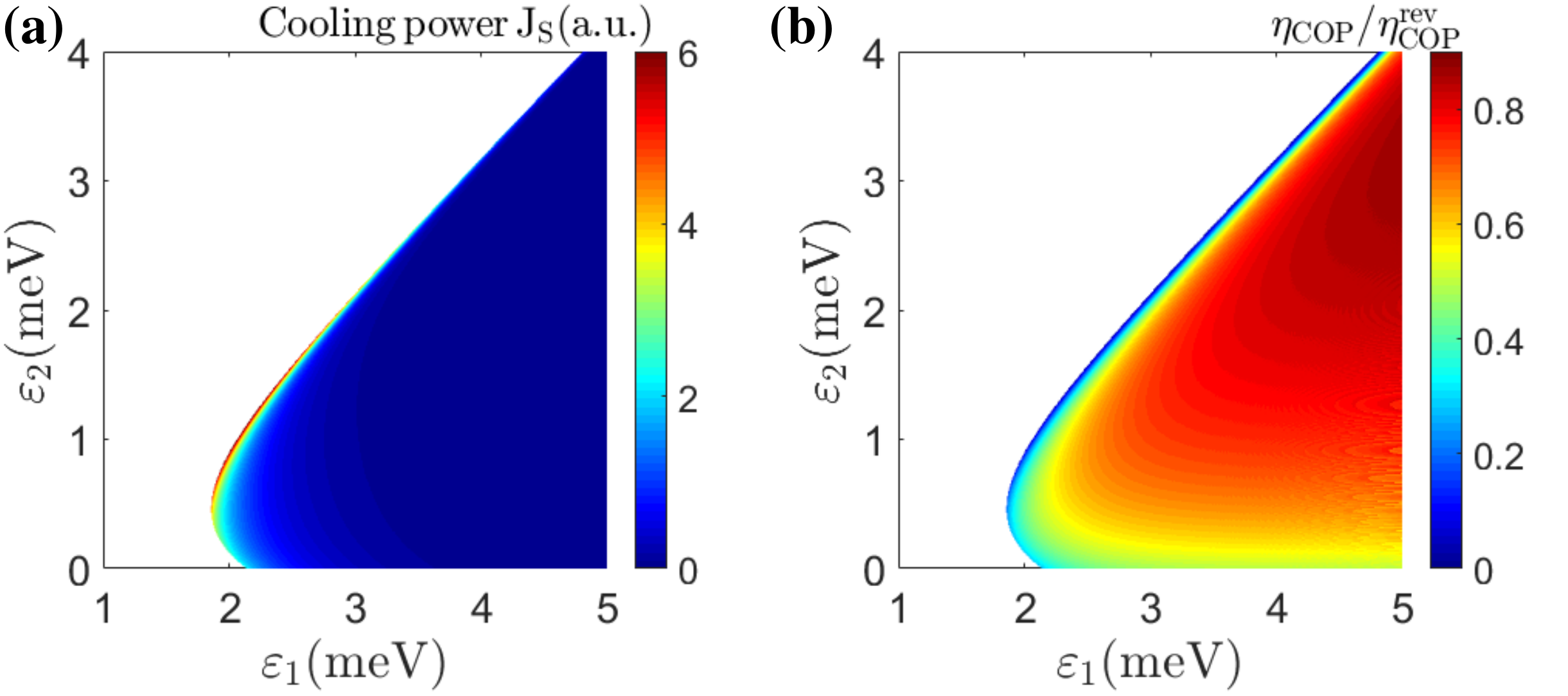}
\caption{(a) Cooling power $J_S$ and (b) COP ratio $\eta_{\rm COP}/\eta_{\rm COP}^{\rm rev}$ as functions of the QDs energies $E_{u1}=E_{d1}=\vep_1$ and $E_{u2}=E_{d2}=\vep_2$ for $J_{in}=0$. Other parameters: $\mu_S=\mu_D=0$, $k_BT_D=1$~meV, $k_BT_H=6$~meV, $T_C=1/(2/T_D-1/T_H)$, $E_C=1$~meV, and $\Gamma_0=0.1$~meV.}~\label{fig:Jin0}
\end{figure}

In Fig.~\ref{fig:ep2}, we show that in our system it is indeed possible to realize the cooling by thermal current with vanishing total injected heat, $J_{in}=0$. We find that the cooling by thermal current effect emerges in the whole region of $0\le \vep_2\le 5$~meV, regardless of the sign change and the vanishing of the total heat injected into the QDs $u1$ and $d1$ from the two heat baths $H$ and $C$.

Under the condition of Maxwell's demon, $J_{in}=0$, we then begin to study the optimal energy configuration for the QD energies, $E_{u1}=E_{d1}=\vep_1$ and $E_{u2}=E_{d2}=\vep_2$ to obtain the lowest temperature of the source and the heat bath $H$. From Fig.~\ref{fig:minTHTS}(a), we find that the lowest temperature $T_S$ when $\varepsilon_1>\varepsilon_2$ and $\varepsilon_1\ge2$~meV. The white areas represent the parameter regions where the Maxwell's demon effect cannot be achieved, i.e., $J_S<0$. Similarly, we also study the minimum temperature of the heat bath $H$ under the requirement of Maxwell's demon. In Fig.~\ref{fig:minTHTS}(b) we can see that it does not require too high temperature of the heat bath $H$ in order to achieve the Maxwell's demon effect in the region $\varepsilon_1>\varepsilon_2$. Meanwhile, we also find that the optimal parameter region for realization of Maxwell's demon effect is that $\varepsilon_1>\varepsilon_2$ from Figs.~\ref{fig:minTHTS}(a) and \ref{fig:minTHTS}(b).

In Fig.~\ref{fig:Jin0},  we show the cooling power $J_S$ and COP for the cooling by thermal current effect $\eta_{\rm COP}$ as functions of QD energies $\varepsilon_1$ and $\varepsilon_2$ for $J_{in}=0$. In accordance with Figure \ref{fig:minTHTS}, the maximum of the cooling efficiency can be achieved when $\varepsilon_1>\varepsilon_2$ but the value of the power is small. The results are consistent with the trade-off of power and efficiency~\cite{JiangPRE,PED,trade-off}.

\section{conclusion and discussions}
In this work, we have demonstrated a mode of the cooling by thermal current effect, the transverse thermoelectric effect and Maxwell's demon effect, through a four-terminal four QDs thermoelectric device (the four terminals are the source, the drain and two electronic thermal baths, respectively). We have demonstrated that the Coulomb interaction between the two circuits separated by an insulating layer is the key factor and the inelastic thermoelectric transport in the two circuits leading to these two unconventional thermoelectric effects. The source can be cooled by passing a thermal current between the two electronic heat baths.

Besides, we studied the Onsager linear-response relations, the Seebeck efficient and the figure of merits of the four-terminal devices. We investigated the relationships between the power and efficiency in the linear or nonlinear response regime.

In addition, we also showed that the two thermal baths can be exploited for power generation and cooling the low-temperature heat bath (or heating the high-temperature heat bath) in a demonlike way, where the working substance does not exchange any energy or particle with the two thermal baths. Our study may be beneficial for autonomous and energy-efficient electric application on the nanoscale devices.

\section{Acknowledgment}
M.X. and T.Y. acknowledge support from the Science and Technological Fund of Anhui Province for Outstanding Youth (Grant No. 1508085J02), the National Natural Science Foundation of China (Grant No.61475004) and the Chinese Academy of Sciences (Grant No.XDA04030213). R.W. J.L and J.-H.J acknowledge support from the National Natural Science Foundation of China (NSFC Grant No. 11675116, No. 12074281), the Jiangsu distinguished professor funding and a Project Funded by the Priority Academic Program Development of Jiangsu Higher Education Institutions (PAPD). J.L also acknowledges support from the China Postdoctoral Science Foundation (Grant No. 2020M681376).

\bibliography{Ref-demon}

\end{document}